\documentclass[twoside,twocolumn,9pt]{article}
\usepackage{extsizes}
\usepackage[super,sort&compress,comma]{natbib} 
\usepackage[version=3]{mhchem}
\usepackage[left=1.5cm, right=1.5cm, top=1.785cm, bottom=2.0cm]{geometry}
\usepackage{balance}
\usepackage{mathptmx}
\usepackage{sectsty}
\usepackage{comment}
\usepackage{graphicx} 
\usepackage{lastpage}
\usepackage[format=plain,justification=justified,singlelinecheck=false,font={stretch=1.125,small,sf},labelfont=bf,labelsep=space]{caption}
\usepackage{float}
\usepackage{fancyhdr}
\usepackage{fnpos}
\usepackage[english]{babel}
\addto{\captionsenglish}{%
  
}
\usepackage{array}
\usepackage{droidsans}
\usepackage{charter}
\usepackage[T1]{fontenc}
\usepackage[usenames,dvipsnames]{xcolor}
\definecolor{cream}{RGB}{255,253,208}
\usepackage{setspace}
\usepackage[compact]{titlesec}
\usepackage{hyperref}
\usepackage{siunitx}
\usepackage{here}

\usepackage{epstopdf}
\newcommand{\fixred}[1]{{\color{black}#1}}

\begin{document}

\pagestyle{plain}
\thispagestyle{plain}

\makeFNbottom
\makeatletter
\renewcommand\LARGE{\@setfontsize\LARGE{15pt}{17}}
\renewcommand\Large{\@setfontsize\Large{12pt}{14}}
\renewcommand\large{\@setfontsize\large{10pt}{12}}
\renewcommand\footnotesize{\@setfontsize\footnotesize{7pt}{10}}
\makeatother

\renewcommand{\thefootnote}{\fnsymbol{footnote}}
\renewcommand\footnoterule{\vspace*{1pt}%
\color{cream}\hrule width 3.5in height 0.4pt \color{black}\vspace*{5pt}} 
\setcounter{secnumdepth}{5}

\makeatletter 
\renewcommand\@biblabel[1]{#1}            
\renewcommand\@makefntext[1]%
{\noindent\makebox[0pt][r]{\@thefnmark\,}#1}
\makeatother 
\renewcommand{\figurename}{\small{Fig.}~}
\sectionfont{\sffamily\Large}
\subsectionfont{\normalsize}
\subsubsectionfont{\bf}
\setstretch{1.125} 
\setlength{\skip\footins}{0.8cm}
\setlength{\footnotesep}{0.25cm}
\setlength{\jot}{10pt}
\titlespacing*{\section}{0pt}{4pt}{4pt}
\titlespacing*{\subsection}{0pt}{15pt}{1pt}

\fancyhf{}
\renewcommand{\headrulewidth}{0pt}
\renewcommand{\footrulewidth}{0pt}
\setlength{\arrayrulewidth}{1pt}
\setlength{\columnsep}{6.5mm}
\setlength\bibsep{1pt}

\makeatletter 
\newlength{\figrulesep} 
\setlength{\figrulesep}{0.5\textfloatsep} 

\newcommand{\topfigrule}{\vspace*{-1pt}%
\noindent{\color{cream}\rule[-\figrulesep]{\columnwidth}{1.5pt}} }

\newcommand{\botfigrule}{\vspace*{-2pt}%
\noindent{\color{cream}\rule[\figrulesep]{\columnwidth}{1.5pt}} }

\newcommand{\dblfigrule}{\vspace*{-1pt}%
\noindent{\color{cream}\rule[-\figrulesep]{\textwidth}{1.5pt}} }

\makeatother

\twocolumn[
  \begin{@twocolumnfalse}
\sffamily
\begin{tabular}{p{18cm}}
\noindent\LARGE{\textbf{Elastocapillary lifting and encapsulation of water by a triangular elastic film under gravity}} \\
\vspace{0.3cm} \\
\noindent\large{Kyoko Shibata,\textit{$^{a\dag}$} Hana Kanda,\textit{$^{b\dag}$}, Yoshimi Tanaka\textit{$^{b}$} and Yutaka Sumino\textit{$^{a,c,d,*}$}} \\
\vspace{0.3cm} \\
\noindent\normalsize{
We investigate the encapsulation of water by a thin elastic film as a minimal model of elastocapillary self-folding with fluid transport. An equilateral triangular polydimethylsiloxane film is lifted quasi-statically from a water surface, while its side length and thickness are systematically varied. Depending on these parameters, the film exhibits three distinct morphologies: folding, recoiling, and liquid encapsulation. We show that the observed morphology is selected by the competition between surface energy, gravitational energy of the liquid, and bending energy of the film. In particular, encapsulation occurs in a narrow parameter region corresponding to the intersection of the elastocapillary, elastogravity, and capillary length scales. 
This result provides a simple physical criterion for liquid encapsulation by elastic films, based on the balance of bending, capillary, and gravitational energies.
} \\

\end{tabular}

 \end{@twocolumnfalse} \vspace{0.6cm}

  ]

\renewcommand*\rmdefault{bch}\normalfont\upshape
\rmfamily
\section*{}
\vspace{-1cm}


\footnotetext{\textit{$^{a}$Department of Applied Physics, Tokyo University of Science, 6-3-1 Nijuku, Katsushika-ku, Tokyo 125-8585, Japan.}}
\footnotetext{\textit{$^{b}$Graduate School of Innovative and Practical Studies, Yokohama National University, Yokohama 240-8501, Japan.}}, \footnotetext{\textit{$^{c}$WaTUS and DCIS, Research Institute for Science \& Technology, Tokyo University of Science, 6-3-1 Nijuku, Katsushika-ku, Tokyo 125-8585, Japan.}}
\footnotetext{\textit{$^{d}$Faculty of Engineering and Physical Sciences, University of Surrey, Guildford, Surrey GU2 7XH, United Kingdom.}}
\footnotetext{$^*$E-mail: ysumino@rs.tus.ac.jp}


\footnotetext{\dag~These authors contributed equally to this work.}


\section{Introduction}
Interfacial tension dominates gravitational effects when the characteristic length of a system becomes smaller than the capillary length, $L_c=\sqrt{\gamma/(\rho g)}$. Under such conditions, capillary forces can drive counterintuitive fluid motions, such as capillary rise and the spontaneous climbing of droplets against gravity~\cite{Washburn1921, Extrand2012, Chaudhury1992, Sumino2005b}. These phenomena originate from the tendency of interfacial tension to minimize surface area, leading to cohesive interactions between fluids and surrounding structures. When the characteristic length scale is sufficiently small, this balance between surface energy and gravity provides a simple physical framework for understanding a wide range of capillarity-driven behaviors.

\begin{figure*}
  \centering
  \includegraphics{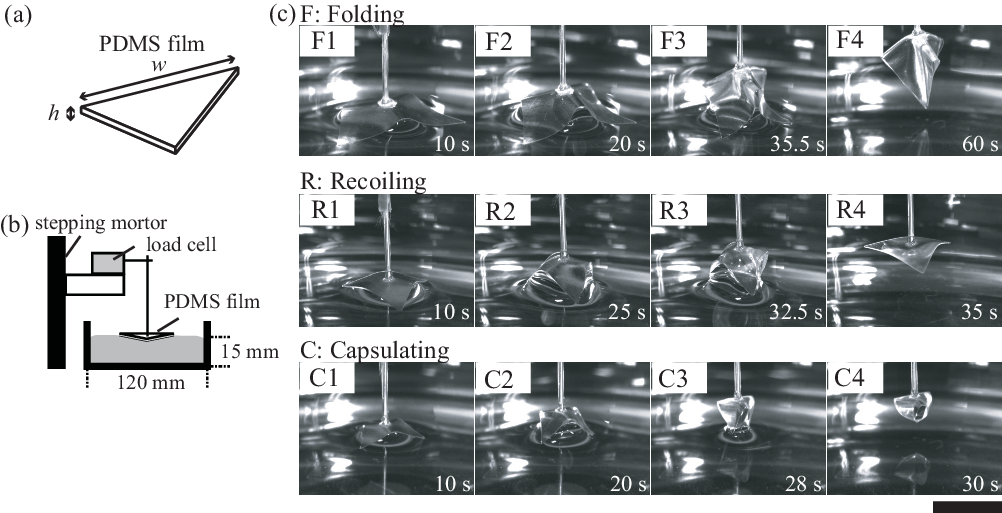}
  \caption{\label{fig:experimentalsystem}
 (a) Geometry of the PDMS film. The film thickness $h$ and side length $w$ were systematically varied. (b) Schematics of experimental setup. The PDMS film was lifted vertically from the water surface by a wire attached at its center. (c) Representative snapshots showing three distinct deformation modes of the film: folding (F; $h$ = 98 {\textmu}m, $w$ = 20 mm), recoiling (R; $h$ = 132.25 {\textmu}m, $w$ = 12.5 mm), and capsulating (C; $h$ = 98 {\textmu}m, $w$ = 10 mm). The time $t=0$ corresponds to the onset of lifting. Scale bar: 10 mm.}
\end{figure*}

When solid boundaries interacting with a liquid are soft and deformable, interfacial tension can induce significant elastic deformation, leading to elastocapillary phenomena. A droplet on an elastomer creates ridges at its contact line, thereby modifying its wetting dynamics~\cite{style2017elastocapillarity}. Similarly, thin elastic filaments can undergo spontaneous twisting driven by the draining of liquids~\cite{kovanko2019capillary}, and micro to nanoscale resist structures may collapse during drying due to capillary forces~\cite{Tanaka1993}. These examples demonstrate that elastocapillary effects not only alter local wetting behavior but can trigger large-scale deformations of soft solids. Such deformation of soft boundaries, in turn, can strongly influence the dynamics of interacting fluids. Beyond engineered systems, similar elastocapillary and capillarity-driven mechanisms are widely exploited in nature, for example, in the water-collecting structures of plants~\cite{bhushan2019bioinspired,shimamura2016marchantia, ju2012multi}, the locomotion of water-walking insects~\cite{Hu2005,Hu2003, XuefengGao2004}, and emerging soft robotic systems~\cite{Yoshii2016, Zhu2021}.

Exploiting the coupling between elasticity and capillarity, capillary origami enables the spontaneous formation of three-dimensional structures from two-dimensional elastic films~\cite{py2007capillary}. 
By appropriately designing the film shape, such self-folding can be harnessed to capture and lift liquid droplets~\cite{reis2010grabbing}. In their study, the folding behavior and resulting functions were primarily controlled by modifying the film geometry, for example, through the number of petals. Related approaches have also demonstrated elastomer-based devices capable of liquid pipetting driven by capillary forces~\cite{nakamura2018plant}.

Despite these advances, the physical criteria governing the onset of liquid encapsulation and self-folding remain unclear, particularly in relation to the competing roles of elasticity, surface tension, and gravity. In this study, we address this issue by fixing the film shape to the simplest symmetric geometry--an equilateral triangle--and systematically varying its side length and thickness. This minimal design allows us to isolate the essential parameters controlling elastocapillary self-folding and liquid encapsulation.
This approach enables us to construct a phase diagram for the encapsulation behavior and to interpret it using characteristic length scales associated with bending, capillarity, and gravity.

In the following sections, we describe an experimental system in which an equilateral triangular elastic film is lifted quasi-statically from a water surface while its side length and thickness are systematically varied. Depending on these parameters, the film exhibits three distinct deformation modes: folding, recoiling, and liquid encapsulation. We analyze these behaviors by considering the competition between surface energy, gravitational energy of the liquid, and bending energy of the film.

\section{Experimental setup}
\subsection{Sample preparation}
Elastic films with thickness $h$= 38--160 \SI{}{\micro \meter} made of polydimethylsiloxane (PDMS) were prepared following standard procedures. The base and curing agents of SILPOT 184 (Dow Chemical) were mixed at a weight ratio of 10:1. The mixture was spin-coated onto a PTFE substrate at rotation speeds ranging from 500 to 1500 rpm to form thin films. The films were then cured at 100 \SI{}{\degreeCelsius} for 1 h to obtain elastic PDMS sheets. After curing, the films were cut into equilateral triangular shapes with a side length $w$, as shown in Fig.~\ref{fig:experimentalsystem}(a). The Young's modulus of the PDMS film was taken as $E$ = 0.7 $\times 10^{6}$ \SI{}{\pascal}~\cite{Fitzgerald2019}. 

The liquid used in this study was pure water prepared using a Millipore Milli-Q system. The surface tension $\gamma$, density $\rho$, and viscosity $\mu$ of water were taken as 72.0 \SI{}{\milli \newton \per \meter}, 1.0 $\times 10^{3}$ \SI{}{\kilogram \per \cubic \meter}, and 1.0 $\times 10^{-3}$ \SI{}{\pascal \second}, respectively.

\subsection{Experimental apparatus}
The schematic experimental setup is shown in Fig.~\ref{fig:experimentalsystem}(b). Pure water was placed in a Petri dish with a diameter of 120 \SI{}{\milli \meter} and a depth of 15 \SI{}{\milli \meter}. The elastic film was attached at its center to an iron wire using cyanoacrylate adhesive (Aron Alpha, Toagousei Co., Itd.).
The wire was connected to a load cell (LVS-5VGA, Kyowa Electronic Instruments Co., Ltd.) mounted on a vertical translation stage driven by a stepping motor (SGSP20-35, Sigma Koki Co., ltd.).

The film was initially positioned at the water surface and then lifted vertically at a constant speed of 1.0 \SI{}{\milli \meter \per \second}.
The deformation of the film during lifting was recorded using a CMOS camera (DMK37BUX273, The Imaging Source) at 20 \SI{}{\hertz}. The force measured by the load cell was recorded using a data logger (GL7000, Graphtec Co.) and used to evaluate the weight $mg$ of the lifted water.

\section{Results}
\begin{figure}[tb]
\centering
\includegraphics{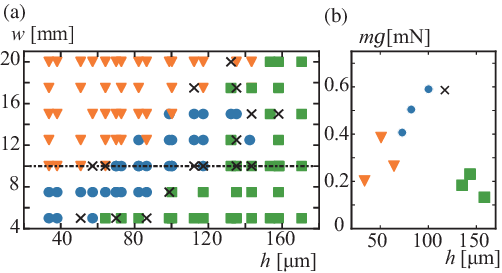}
\caption {\label{fig:phases} 
(a) Phase diagram of the deformation modes of the lifted film as a function of film thickness $h$ and side length $w$. 
Green squares, orange triangles and blue circles represent folding, recoiling, and capsulating modes, respectively [see Fig.~\ref{fig:experimentalsystem}(c)]. The deformation modes were classified based on the final configuration of the film observed during the lifting process. The dashed line indicates the parameter set used in panel (c). All experimental data points are plotted in the phase diagram.
(b) Weight of water, $mg$, lifted by the film as a function of film thickness $h$, measured for a fixed side length $w$ = 10 \SI{}{\milli \meter}. 
}
\end{figure}

\begin{figure}[tb]
\centering
\includegraphics{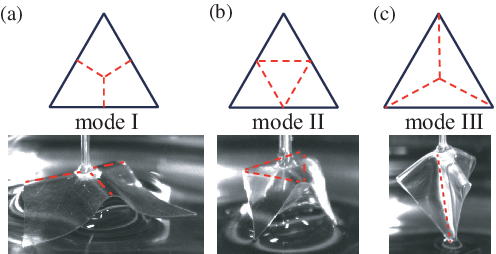}
\caption {\label{fig:modes} 
Characteristic deformation modes of the elastic film, classified as (a) Mode I, (b) Mode II, and (c) Mode III. The dashed lines indicate regions of large mean curvature. Bottom images show the corresponding regions of high mean curvature in the experimental snapshots shown in Fig.~\ref{fig:experimentalsystem}(c).
}
\end{figure}

When an elastic film was lifted from the surface of a liquid, distinct deformation behaviors were observed depending on the film thickness and side length. As shown in Fig.~\ref{fig:experimentalsystem}(c), the film exhibited three characteristic deformation modes: folding, recoiling, and liquid encapsulation. In the following, we describe these modes based on representative time sequences and then summarize the observed behaviors in a phase diagram.

\subsection{Deformation modes of elastic films}
As shown in these snapshots, the film initially lifted liquid due to wetting of the PDMS film surface in all cases. When the film was sufficiently thin and flexible, it folded spontaneously with threefold symmetry around the attached wire while the liquid was gradually drained away [Fig.~\ref{fig:experimentalsystem}(c:F1-4)]. In contrast, when the film was sufficiently rigid to resist deformation, it recoiled back to a nearly flat shape, retaining only a small liquid droplet attached at its bottom [Fig.~\ref{fig:experimentalsystem}(c:R1-4)]. In an intermediate parameter range between these two behaviors, liquid encapsulation was observed, in which the liquid was spontaneously wrapped inside the film [Fig.~\ref{fig:experimentalsystem}(c:C1-4)].

\subsection{Phase diagram of film morphologies}
Fig.~\ref{fig:phases}(a) shows the phase diagram of the observed film morphologies as a function of film thickness $h$ and side length $w$.  The deformation modes were classified based on the final film configuration. Three distinct regions corresponding to folding, recoiling, and capsulating modes are identified. The capsulating mode appears in an intermediate region between the folding and recoiling regimes. A small number of marginal cases, where mixed or asymmetric behaviors
were observed, are also indicated.

The occurrence of liquid encapsulation is further quantified by the weight of the lifted liquid, $mg$, as shown in Fig.~\ref{fig:phases}(b). These data were obtained for films with a fixed side length $w$=10 mm. Larger values of $mg$ were observed when the film exhibits the capsulating mode. As the system transitions from folding to capsulating, the lifted liquid weight increases gradually. In contrast, a sharp decrease in $mg$ is observed when the system transitions from capsulating to recoiling at larger $h$. 

In addition to these dominant behaviors, marginal cases were observed, including mixed capsulating-recoiling states, where a deformed plate retains a finite liquid droplet at its bottom, as well as asymmetric folding without threefold symmetry. These marginal events are indicated by black crosses in Fig.~\ref{fig:phases} (a) and (b).

The lifting speed was varied from 0.1 to 20~\SI{}{\milli\meter\per\second}, and no significant
difference in the observed deformation modes was found compared to the
standard condition of 1~\SI{}{\milli\meter\per\second}. Over this range of speeds, the capillary number $Ca=\mu V/\gamma$ varied from $10^{-6}$ to $10^{-4}$, indicating that the viscous effect was negligible. The Reynolds number $Re=\rho \mu V/L$, was of the order unity or smaller in the present system. These observations indicate that the deformation modes and phase behavior reported here are insensitive to the lifting speed within the accessible experimental range.

\section{Discussion}
\subsection{Deformation modes of elastic films}

The deformation behaviors observed in the lifting process can be interpreted in terms of three characteristic deformation modes (Mode I–III), as illustrated in Fig.~\ref{fig:modes}(a-c). These modes preserve the threefold symmetry of the equilateral triangular film and are distinguished by the locations of regions of high curvature.

In the folding regime, the deformation is initially dominated by Mode I [Fig.~\ref{fig:modes}(a)], where ridges form along lines connecting the lifting point to the midpoints of the film edges. Geometrical constraints at the vertices due to capillary force at the air-water interface prevent vertical deformation, leading to the coexistence of Mode I and Mode III [Fig.~\ref{fig:modes}(c)] at later stages. This combined deformation ultimately results in expulsion of the liquid.

In contrast, recoiling behavior is associated with Mode II [Fig.~\ref{fig:modes}(b)], in which the regions near the vertices bend downward. As the film is lifted, capillary breaking occurs, and the elastic energy stored in the film drives the system back to a nearly flat configuration.

Liquid encapsulation emerges in an intermediate regime, where the deformation evolves from a mixed Mode I–II configuration toward a Mode II–dominated state. In this case, the approach of the three vertices facilitates the formation of a transient liquid bridge, which subsequently undergoes capillary breaking to form a closed capsule.

\subsection{Characteristic length scales and energy competition}
The phase behavior observed in the present experiments can be interpreted in terms of characteristic length scales, namely the elastogravity length $L_{eg}$~\cite{reis2010grabbing} and the elastocapillary length $L_{ec}$~\cite{reis2010grabbing}.
These characteristic lengths are given by
\begin{align}\label{Eq:characteristic_length}
L_{eg}=\left(\frac{B}{\rho g}\right)^{1/4}, \,\,L_{ec}=\left(\frac{B}{\gamma}\right)^{1/2}
\end{align}
These characteristic lengths provide natural criteria for comparing the system size $L$ with the relevant elastocapillary and elastogravity scales.

\begin{table}
\centering
\caption{\label{table:variables} Variables used for dimensional analysis}
\begin{tabular}{ccc}
\hline
Variable & description & values \\
\hline
$E$ & Young's modulus & 0.7 $\times$ 10$^6$~\SI{}{\pascal} \\
$\nu$ & Poisson's ratio & 0.5 \\
$\rho$ &density of water &  1.0 $\times$ 10$^3$~\SI{}{\kilogram \per \cubic \meter } \\
$g$ & Gravitational acceleration & 10~\SI{}{\meter \per \square \second } \\
\hline
\end{tabular}
\end{table}

In the case when $L \gg L_{eg}$, the bending energy becomes subdominant compared to the gravitational energy, and the film tends to undergo folding with significant out-of-plane deformation when the film is lifted. Indeed, $L_{eg}$ is known to correspond to the typical wavelength of the elastic film under the effect of bottom liquids~\cite{reis2010grabbing}. A similar argument applies to the present system. On the other hand, when $L \ll L_{eg}$, the bending energy is dominant compared to the gravitational energy, and the film tends to remain flat and recoil back.

Similarly, when $L \gg L_{ec}$, the bending energy is subdominant compared to the surface energy, and the film tends to fold into a racket-like shape~\cite{py2007capillary}. When $L \ll L_{ec}$, capillary forces are insufficient to bend the film, and encapsulation of water does not occur

Interestingly, when $L \simeq L_{ec}$ and $L \simeq L_{eg}$, the capillary length $L_{cg} = (\gamma/\rho g)^{1/2}$ also becomes comparable to $L$. Therefore, the intersection of these characteristic length scales provides a natural condition for liquid encapsulation. This suggests that encapsulation emerges when bending, capillary, and gravitational energies become comparable.

\begin{figure}
\centering
\includegraphics[width=8.8cm]{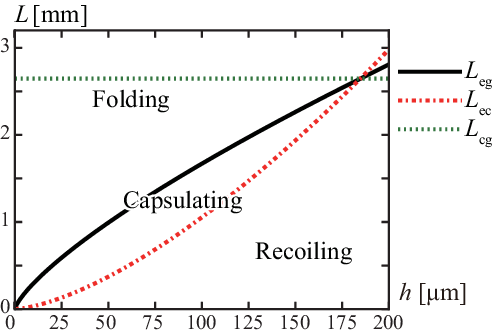}
\caption {\label{fig5} 
    The characteristic length of the system, elastogravity, $L_{eg}$ (black solid line), elastocapillary, $L_{ec}$ (red dash-dotted line), and capillary length $L_{cg}$ (green dotted line). Each line is drawn based on Eq.~\eqref{Eq:characteristic_length}. 
}
\end{figure}

We note that $B = Eh^3/\{12(1-\nu^2)\}$, where $B$ is proportional to $h^3$. Thus, $L_{eg} \sim h^{3/4}$ and $L_{ec} \sim h^{3/2}$. Using the parameters listed in Table~\ref{table:variables}, these characteristic lengths are plotted in Fig.~\ref{fig5}. When plotted as a function of the film thickness $h$, $L_{eg}$ increases as $h^{3/4}$ while $L_{ec}$ increases as $h^{3/2}$. As a result, the two characteristic lengths inevitably intersect for some range of $h$, irrespective of the precise numerical prefactors.
\fixred{This intersected region is consistent with the parameter range where water encapsulation was observed in the experiments, as seen in the phase diagram shown in Fig.~\ref{fig:phases}(a).}

\section{Summary and conclusion}

In this study, we created an experimental system in which a droplet of water is lifted by an elastic film, which is attached to a wire that is moved upward at a constant speed. We controlled the film's side length and thickness while keeping the water's surface tension constant. It was found that water can be lifted up in a limited parameter region in which water was encapsulated by the film (capsulating). Outside the capsulating regime, two other deformation behaviors were observed. In one region, the film folded while retaining only a small amount of water (folding). In the other region, the film exhibited recoiling dynamics and ultimately returned to a nearly undeformed state with a small amount of water (recoiling).

With a simple assumption on the geometry, we confirmed that the film's encapsulating geometry can be seen in the corresponding parameter region in terms of characteristic length scale, elastogravity, elastocapillary, and capillary length. Surprisingly, despite the simplified estimation of these energies, the obtained phase diagram showed semi-quantitative agreement with the experiment.

In this study, we kept the film's uplift speed constant at 1.0 mm/s. Our preliminary trial showed that the dynamic aspect should be considered when the depth of the liquid reservoir is as small as 1 mm and the liquid viscosity is 1 \SI{}{\pascal \second} so that the effective capillary number for the experiment is of the order of 1. In such a situation, the dynamic effect should be interesting to consider. We leave such a situation for future study.

Our results show that simple elastic films can be used to manipulate liquids by appropriately choosing the thickness and side length. For potential application, it would be interesting to consider the effect of handling non-Newtonian liquids with a film for future study. By proceeding in this direction, we may even use this combination to transport liquids and gels.

\section*{Author contributions}
Y.T. and Y.S. conceived the experimental design and supervised this study. All experiments were conducted and analyzed by K.S., K.H., Y.T. and Y.S. 

\section*{Conflicts of interest}
There are no conflicts to declare.

\section*{Data availability}
The authors confirm that the data supporting the findings of this study are available within the article.


\section*{Acknowledgements}
This work is partially supported by the Grant-in-Aid for Japan Society for Promotion of Science, KAKENHI (Grants No.\ JP16H06478 , No.\ 21H01004 and No.\ 21H01006).  This work was also supported by JSPS and PAN under the Japan-Poland Research Cooperative Program ``Spatio-temporal patterns of elements driven by self-generated, geometrically constrained flows'', the JSPS Core-to-Core Program ``Advanced core-to-core network for the physics of self-organizing active matter'' (JPJSCCA20230002)



\balance


\providecommand*{\mcitethebibliography}{\thebibliography}
\csname @ifundefined\endcsname{endmcitethebibliography}
{\let\endmcitethebibliography\endthebibliography}{}

\bibliographystyle{rsc} 

\end{document}